\documentclass[3p,10pt,a4paper,twoside,fleqn,sort&compress]{elsarticle}
\makeatletter
\def\ps@pprintTitle{%
  \let\@oddhead\@empty
  \let\@evenhead\@empty
  \let\@oddfoot\@empty
  \let\@evenfoot\@oddfoot
} \makeatother

\usepackage{graphicx}
\usepackage{verbatim}
\usepackage{amsmath}
\usepackage[varg]{pxfonts}
\usepackage{eepic}
\usepackage{amssymb}

\usepackage{mathtools}
\usepackage[all,cmtip]{xy}

\usepackage{tikz}
\usetikzlibrary{matrix,arrows}

\newtheorem{theorem}{Theorem}[section]

\newtheorem{remark}[theorem]{Remark}

\newcommand{\mc}{\mathcal}
\newcommand{\comp}{\circ}

\newcommand{\mcP}{P}
\newcommand{\mcU}{U}
\newcommand{\mcQ}{Q}
\newcommand{\mcPQ}{PQ}

\newcommand{\Hv}{\mc N}

\newcommand{\isoXtoR}{h^{-1}}
\newcommand{\isoRtoX}{h}

\newcommand{\q}{q}

\newcommand{\op}{\oplus}
\newcommand{\cop}{\tau}

\newcommand{\ec}{\mc I}

\newcommand{\ra}{\rightarrow}

\newcommand{\sR}{\mathbb R}
\newcommand{\sN}{\mathbb N}

\newcommand{\pvec}{(p_1, \dots, p_n)}
\newcommand{\qvec}{(q_1, \dots, q_m)}

\newcommand{\escp}[1]{p_{#1}^{(q)}}
\newcommand{\escr}[1]{r_{#1}^{(q)}}
\newcommand{\escq}[1]{q_{#1}^{(q)}}

\newcommand{\Hsm}{\mc D}

\newcommand{\oplam}{\op_q}

\newcommand{\lamg}{(1-q)}

\newcommand{\jp}{}

\newcommand{\Hjiz}[1]{\widetilde{\mc D}_{#1}}
\begin{document}

\begin{frontmatter}

\title{\Large\bf Comments on\\ "On q-non-extensive statistics with non-Tsallisian entropy"}

\author[misanu]{Velimir M. Ili\'c\corref{cor}}
\ead{velimir.ilic@gmail.com}

\author[fosun]{Miomir S. Stankovi\'c}
\ead{miomir.stankovic@gmail.com}

\cortext[cor]{Corresponding author. Tel.: +38118224492; fax:
+38118533014.}
\address[misanu]{Mathematical Institute of the Serbian Academy of Sciences and Arts, Kneza Mihaila 36, 11000 Beograd, Serbia}
\address[fosun]{University of Ni\v s, Faculty of Occupational Safety, \v Carnojevi\'ca 10a, 18000 Ni\v s, Serbia}

\begin{abstract}
\begin{minipage}{15.5cm}

Recently, in 
[P. Jizba and J. Korbel, Physica A 444, 2016, 808–827], four
generalized Shannon-Khinchin [GSK] axioms have been proposed and a
generalized entropy which uniquely satisfies the GSK axioms has
been derived. In this comment, we show that the unique class of
the entropies derived in the aforementioned paper is not correct,
as it violates the fourth GSK axiom, and we derive the correct
one. Nevertheless, the class of entropies proposed in the
commented paper still can serve as a basis for generalized
statistical mechanics. We propose a new axiomatic system which
characterizes the class of entropies.\\

\end{minipage}
\end{abstract}

\end{frontmatter}
%
\section*{}

\newcommand{\Hs}{\mc S}
\newcommand{\sRp}{\sR^+_0}

\section{Preliminary notions}

The set of positive real numbers will be denoted with $\sR^+$ and
the set of nonnegative real numbers will be denoted with $\sRp$.
The set of all $n$-dimensional distributions will be denoted with
\begin{equation}
   \Delta_n \equiv \left\{ (p_1, \dots , p_n) \Big\vert \; p_k \in \sRp,
      \sum_{k=1}^{n} p_k = 1 \right \};\quad  n>1.
  \label{Delta}
\end{equation}
For $\mcP = \pvec \in \Delta_n$ and $\mcQ = \qvec \in \Delta_m$, a
\textit{direct product}, $\mcP \star \mcQ \in \Delta_{nm}$ is
defined as
\begin{equation}
\mcP \star \mcQ = (p_1q_1, p_1q_2, \dots, p_n q_m).
\end{equation}
%
Let in the following $q\in \sR^+$. For a distribution $\mcP \in
\Delta_n$, we define a \emph{$q$-escort distribution}
$\mcP^{(q)}\in \Delta_n$ with
\begin{equation}
\mcP^{(q)}=(\escp{1},\dots,\escp{n}); \quad\quad
\escp{k}=\frac{p_k^q}{\sum_{i=1}^n p_i^q},\quad k=1,\dots,n.
\end{equation}
A $\oplam$-addition \cite{Nivanen_Wang__03} is defined with
\begin{equation}
\label{SM: op} u \oplam v = u + v +\lamg u v; \quad u, v \in \sR.
\end{equation}
Let us define an increasing, continuous, and invertible function
$\isoRtoX: \sR \ra \sR$ and the inverse mapping $\isoXtoR: \sR \ra
\sR$ as
\begin{equation}
\label{sec: sm: h}
\isoRtoX(x) = %
\begin{dcases}
\quad\quad x, &\mbox{ for }\quad q = 1 \\
\frac{2^{(1-q) \jp x} - 1}{1-q}, \quad 
&\mbox{ for }\quad q \neq 1
\end{dcases}
\quad\quad\Leftrightarrow\quad\quad
\isoRtoX^{-1}(x) = %
\begin{dcases}
\quad\quad x, &\mbox{ for }\quad q = 1 \\
\frac{1}{1-q} \log_2((1-q) x +1), \quad 
&\mbox{ for }\quad q \neq 1
\end{dcases}.
\end{equation}
Then, the following equality holds:
\begin{equation}
\label{sec: tsallis: op definition}%
\isoRtoX(x+y) = \isoRtoX(x) \oplam \isoRtoX(y), \quad\quad
\isoXtoR(x\oplam y) = \isoXtoR(x)
+ \isoXtoR(y)  
; \quad\quad x, y \in \sR.
\end{equation}

\section{On the generalized Shannon-Khinchin axioms from \cite{Jizba_2016} and a corresponding unique class of entropies}

A generalized entropy is characterized as a unique function $\mc
D_n: \Delta_n \ra \sRp$, which for all $n\in\sN$, $n>1$ satisfies
the following generalization of the Shannon-Khinchin axioms
\cite{Khinchin_57}:

\begin{description}

\item{[A1]} $\mc D_n$ is continuous in $\Delta_n$;

\item{[A2]} $\mc D_n$ takes its largest value for the uniform
distribution, $\mcU_n=\left({1}/{n}, \dots, {1}/{n}\right) \in
\Delta_n$, i.e. 
$\mc D_n(\mcP)\le \mc D_n(\mcU_n)$, for any $\mcP \in \Delta_n$;

\item{[A3]} $\mc D_n$ is expandable: $\mc D_{n+1}(p_1,p_2, \ldots , p_n, 0 ) = \mc D_n(p_1,p_2, \ldots ,
p_n)$ for all $(p_1,\dots,p_n)\in\Delta_n$;

\item{[A4]} Let $\mcP = (p_1, \dots, p_n) \in \Delta_n$, 
$\mcPQ = (r_{11}, r_{12}, \dots, r_{nm}) \in \Delta_{nm}$, $n, m
\in \sN$, $n>1$ such that $p_i = \sum_{j=1}^m r_{i j}$, and
$\mcQ_{ | i} = (q_{1|i}, \dots, q_{m|i}) \in \Delta_m$, where
$q_{j|i} = r_{ij}/p_i$ and $\q \in \sR_+$ is some fixed parameter.
Then,
\begin{equation}
\label{ax 4} \Hsm_{nm}(\mcPQ) = \Hsm_n(\mcP) \oplam \Hsm_m(\mcQ|
\mcP), \quad\text{where}\quad \Hsm_m(\mcQ| \mcP) = f_h^{-1}
\left(\sum_{i=1}^n \escp{i} f_h(\Hsm_m(\mcQ_{ | i})) \right),
\end{equation}
where $f_h$ is an invertible continuous function.

\end{description}

The axiomatic system [A1]-[A4] was first presented in
\cite{Jizba_2004} and latter on in \cite{Jizba_2016}, where a
class of entropies is proposed as the unique one which satisfies
the axioms, as it is summarized in Theorem \ref{theo: jizba}.
However, the proposed unique solution of the axiomatic system is
not correct, as it violates the axiom [A4]. We comment on this in
Remark \ref{rem: jizba incor }. After that, in Theorem \ref{theo:
ilic_stankovic}, we derive the correct unique solution of
[A1]-[A4].

\begin{theorem}\label{theo: jizba} (Incorrect, from Appendix in \cite{Jizba_2016}) The class of entropies which uniquely satisfies [A1]-[A4] is given
by
\begin{align}
\label{entropy: jizba} \widetilde{\mc D}_n(\mcP) = 
\widetilde{\mc D}_n(\mcP)=h\left(-\sum_{k=1}^{n} \escp{k}  \log_2
p_k\right)=
\begin{dcases}
-\sum_{k=1}^{n} p_k  \log_2 p_k, \quad\quad\quad \quad &\mbox{for}
\quad q =1
\\
\frac{1}{1-q} \left(\prod_{k=1}^{n} p_k^{(q-1) \escp{q}}-1
\right)&\mbox{for}\quad q \neq 1.
%
\end{dcases}
\end{align}
\end{theorem}

\begin{remark}\label{rem: jizba incor } (Incorrectness of Theorem \ref{theo:
jizba}) \rm In the proof of the theorem from 
\cite{Jizba_2016} it is (incorrectly) assumed that $\mc D_n(P)$
has the form
\begin{equation}
\label{hjiz2} \Hjiz{n}(\mcP) = f_h^{-1} \left( \sum_{k=1}^{n}
\escp{k} f_h\left(h(\cop \log_2 p_k)\right) \right), \quad \cop
<0,
\end{equation}
and, as a consequence, $f_h$ is found as a linear function of
$h^{-1}$, which leads to the class of entropies \eqref{entropy:
jizba}, which violates [A4].
To see this, let us rewrite [A4] as
\begin{equation}
\label{h a4 wrong} h\left(-\sum_{i=1}^{n} \sum_{j=1}^{m} \escr{ij}
\log_2
r_{ij}\right)=%
h\left(-\sum_{i=1}^{n} \escp{i}  \log_2 p_i\right)\oplam%
h\left(-\sum_{i=1}^{n} \escp{i} \sum_{j=1}^{m} \escq{j|i}  \log_2
q_{j|i}  \right).
\end{equation}
After applying the mapping $h^{-1}$ to the equation \eqref{h a4
wrong}, and according to \eqref{sec: tsallis: op definition}, we
obtain
\begin{equation}
\label{h a4 transformed} -\sum_{i=1}^{n} \sum_{j=1}^{m} \escr{ij}
\log_2
r_{ij}=%
-\sum_{i=1}^{n} \escp{i}  \log_2 p_i - %
\sum_{i=1}^{n} \escp{i} \sum_{j=1}^{m} \escq{j|i}  \log_2
q_{j|i}.
\end{equation}
The equation \eqref{h a4 transformed} 
does not hold, in general, for every $q$. To check this, let $q=2$
and
\renewcommand\arraystretch{1.7}
\begin{equation}
P= ( p_1 , p_2 ) =\left(\frac{1}{2},\frac{1}{2}\right) \quad
\text{and} \quad R=
\begin{pmatrix}
r_{11} & r_{12} \\
r_{21} & r_{22}
\end{pmatrix}
=
\begin{pmatrix}
\frac{1}{4} & \frac{1}{4} \\
\frac{1}{2} & 0
\end{pmatrix},
\quad \text{so that}
 \quad
\begin{array}{l}
Q_{|1}=\left(q_{1|1}, q_{2|1}\right)= \left(\frac{1}{2}, \frac{1}{2}\right)  \\
Q_{|2}=\left(q_{1|2}, q_{2|2}\right)= \left(1, 0\right)
\end{array},
\end{equation}
where we have used $q_{j|i}=r_{ij}/p_i$. Note that if
$V=(1/2,1/2)$ or $V=(1,0)$, then $V$ is equal to its escort
distribution, $V=V^{(q)}$, so that $P^{(q)}= P$, $Q_{|1}^{(q)}=
Q_{|1}$ and $Q_{|2}^{(q)}= Q_{|2}$ and
\begin{equation}
(p_1^{(q)}, p_2^{(q)}) =
\left(\frac{1}{2}, \frac{1}{2} \right),
\quad \quad%
\begin{pmatrix}
r_{11}^{(q)} & r_{12}^{(q)} \\
r_{21}^{(q)} & r_{22}^{(q)}
\end{pmatrix}
=
\begin{pmatrix}
\frac{1}{2+2^q} & \frac{1}{2+2^q}\\
\frac{2^q}{2+2^q}  & 0
\end{pmatrix}=
\begin{pmatrix}
\frac{1}{6} & \frac{1}{6} \\
\frac{2}{3} & 0
\end{pmatrix},
\quad \quad
\begin{array}{l}
\left(q_{1|1}^{(q)}, q_{2|1}^{(q)}\right)= \left(\frac{1}{2}, \frac{1}{2}\right)  \\
\left(q_{1|2}^{(q)}, q_{2|2}^{(q)}\right)= \left(1,
0\right)
\end{array}
\end{equation}
Therefore, we have:
\begin{equation}
-\sum_{i=1}^{n} \sum_{j=1}^{m} \escr{ij} \log_2
r_{ij}=-\frac{1}{6} \log_2 \frac{1}{4} -\frac{1}{6} \log_2
\frac{1}{4} - \frac{2}{3} \log_2 \frac{1}{2} = \frac{4}{3}
\end{equation}
and
\begin{equation}
-\sum_{i=1}^{n} \escp{i}  \log_2 p_i - %
\sum_{i=1}^{n} \escp{i} \sum_{j=1}^{m} \escq{j|i}  \log_2
q_{j|i}= 1+\frac{1}{2}\cdot 1+ \frac{1}{2}\cdot 0=\frac{3}{2}, 
\end{equation}
and the equality \eqref{h a4 transformed} does not hold, which
means that entropy form from \cite{Jizba_2016} violates the axiom
[A4].
\end{remark}

In the following theorem, we fix the mistake by determining the
correct unique class of functions which satisfies [A1]-[A4], and
the class of functions which $f_h$ belongs to.

\begin{theorem}\label{theo: ilic_stankovic} (Correction to Theorem \ref{theo: jizba}) The class of entropies which uniquely satisfies [A1]-[A4] is given
by the following class:
\begin{align}
\label{gen entr definition} \mc D_n(\mcP) =
\begin{dcases}
\ \cop \jp \sum_{k=1}^{n} p_k  \log_2 p_k, \quad\quad\quad \quad
&\mbox{for} \quad q =1
\\
\frac{1}{1-q} \left( \Big(\sum_{k=1}^{n} p_k^{\q}\Big)^{-\cop}-1
\right),&\mbox{for}\quad q \neq 1
%
\end{dcases}
\quad,\quad\quad \cop<0.
\end{align}
and $f_h$ has the form
: 
\begin{equation}
\label{sec: f_h}
f_h(x) = %
\begin{dcases}
\quad\quad \cop x, &\mbox{ for }\quad q = 1 \\
((1-q) x +1)^{-\frac{1}{\cop}}, \quad 
&\mbox{ for }\quad q \neq 1
\end{dcases} \quad,\quad\quad \cop<0,
\end{equation}
or any linear function of \eqref{sec: f_h}.

\end{theorem}

\textbf{Proof.} Let us define 
$\Hv_n=h^{-1}\comp \mc D_n$ and $f=f_h \comp h$, where $\comp$
denotes the composition of functions. Then, [A1]-[A4] can
equivalently be written as:

\begin{description}

\item{[B1]} $\mc N_n$ is continuous in $\Delta_n$;

\item{[B2]} $\mc N_n$ takes its largest value for the uniform
distribution, $\mcU_n=\left({1}/{n}, \dots, {1}/{n}\right) \in
\Delta_n$, i.e. 
$\mc N_n(\mcP)\le \mc N_n(\mcU_n)$, for any $\mcP \in \Delta_n$;

\item{[B3]} $\mc N_n$ is expandable: $\mc N_{n+1}(p_1,p_2, \ldots , p_n, 0 ) = \mc N_n(p_1,p_2, \ldots ,
p_n)$ for all $(p_1,\dots,p_n)\in\Delta_n$;

\item{[B4]} Let $\mcP = (p_1, \dots, p_n) \in \Delta_n$, 
$\mcPQ = (r_{11}, r_{12}, \dots, r_{nm}) \in \Delta_{nm}$, $n, m
\in \sN$, $n>1$ such that $p_i = \sum_{j=1}^m r_{i j}$, and
$\mcQ_{ |i} = (q_{1|i}, \dots, q_{m|i}) \in \Delta_m$, where
$q_{j|i} = r_{ij}/p_i$ and $\q \in \sRp$ is some fixed parameter.
Then,
\begin{equation}
\mc N_{nm}(\mcPQ) = \mc N_n(\mcP) + \mc N_m(\mcQ| \mcP),
\quad\text{where}\quad \mc N_m(\mcQ| \mcP) = f^{-1}
\left(\sum_{i=1}^n \escp{i} f(\mc N_m(\mcQ_{ | i})) \right),
\end{equation}
where $f$ is an invertible continuous function.

\end{description}
As shown in \cite{Ilic_Stankovic_14}, a function $\mc N_n:
\Delta_n \ra \sRp$ satisfies [B1]-[B4] for all $n>1$, iff it
belongs to the following class:
\begin{align}
\label{Nath entropy: Nath definition} \Hv_n(\mcP) =
\begin{dcases}
\ \cop \jp \sum_{k=1}^{n} p_k  \log_2 p_k, \quad\quad\quad \quad
&\mbox{for} \quad \q =1
\\
\ \frac{\cop}{\q-1} \log_2 \left( \sum_{k=1}^{n} p_k^{\q} \right),
\quad \quad &\mbox{for} \quad \q \neq 1
\end{dcases}\quad,
\quad\quad \cop<0.
\end{align}
Consequently, a function $\mc D_n=h\comp \Hv_n$ satisfies
[A1]-[A4] iff it belongs to the class $\mc D_n=h\left(
\Hv_n(\mcP)\right)$, which is given by \eqref{gen entr
definition}.

The form of the function $f_h$ can be found as follows.
By substituting of the expression \eqref{Nath entropy: Nath
definition} to the axiom [B4],
we can obtain
\begin{equation}
\mc N_m(\mcQ| \mcP)=\mc N_{nm}(\mcPQ) - \mc N_n(\mcP)=
\frac{\cop}{\q-1} \log_2 \sum_{i=1}^{i} p_i^{(\q)} \sum_{j=1}^{m}
q_{j|i}^{\q}=f^{-1} \left(\sum_{i=1}^n \escp{i} f(\mc N_m(\mcQ_{ |
i})) \right),
\end{equation}
where $f$ is the function from the class parameterized by $\q,
\cop \in \sR$: 
\begin{equation}
\label{disc: f(x)} f(x) =
\begin{dcases}
\ \cop \jp x, &\text{for}\quad q = 1 \\
2^{\frac{(q-1)  x}{\cop}}, &\text{for}\quad q \neq 1
\end{dcases}
\quad\quad\Leftrightarrow\quad\quad f^{-1}(x) =
\begin{dcases}
\ \ \frac{ x}{\cop}, &\text{for}\quad\q=1 \\
\frac{\cop\  \log_2{x }}{\q-1}, &\text{for}\quad\q \neq 1.
\end{dcases}
\end{equation}
%
Moreover, $\mc D_m=h \comp \mc N_m$ so that
\begin{equation}
\mc D_m(\mcQ| \mcP)=
f_h^{-1}\left(\sum_{i=1}^n \escp{i}\ f_h^{-1}(\mc D_m(\mcQ_{ |
i})) \right)
=h(\mc N_m(\mcQ| \mcP))=
\left(f\comp h^{-1}\right)^{-1}\left(\sum_{i=1}^n \escp{i}\
f\comp h^{-1}(\mc D_m(\mcQ_{ | i})) \right).%
\end{equation}
Therefore, $f_h$ and $f\comp h^{-1}$ generate the same
quasi-linear mean, and according to a well known result from mean
theory \cite{ Hardy_et_al_34}, $f_h$ is a linear function of
$f\comp h^{-1}$, which proves the theorem. $\square$

\begin{remark}\rm The generalized entropy $\mc D_{n}$ can also be represented as
\begin{equation}
\label{h_gh} \mc D_{n}(\mcP) = g_h^{-1} \left( \sum_{k=1}^{n}
\escp{k} g_h\left(h(\cop \log_2 p_k)\right) \right), \quad \cop
<0,
\end{equation}
where $g_h$ has a form:
\begin{equation}
\label{Reny 1: theorem: g(x)}
g_h(x) = %
\begin{dcases}
\quad\quad \cop x, &\mbox{ for }\quad q = 1 \\
((1-q) x +1)^{\frac{1}{\cop}}, \quad 
&\mbox{ for }\quad q \neq 1,
\end{dcases}
\end{equation}
or any linear function of \eqref{Reny 1: theorem: g(x)}.
Therefore, the assumption \eqref{hjiz2} made in \cite{Jizba_2016},
is valid only in the case when $g_h$ is linear function of $f_h$,
which can is satisfied for $q=1$, i.e. in the case of Shannon
entropy.
\end{remark}

\begin{remark}\rm If an additional normalization axiom is added to
[A1]-[A4]:

\begin{description}

\item{[A5]} $\mc D_n(\frac{1}{2},\frac{1}{2})=h(1)$,

\end{description}
then $\cop=-1$ and the entropy \eqref{gen entr definition} reduces
to Tsallis entropy \cite{Tsallis_88} and the function $f_h$ is a
linear function of $x$. This agrees with the result presented by
Abe in \cite{Abe_00}, who has shown that if $f_h$ is linear, then
[A1]-[A4] uniquely characterize Tsallis entropy. In other words,
if the axiomatic system from \cite{Jizba_2016}, is extended with
the axiom [A5], then it is equivalent to Abe's axiomatic system.

\end{remark}

\begin{remark} \rm The axiomatic system [A1]-[A4] and the corresponding unique class
of entropies can further be generalized if the conditional entropy
in the axiom [A4] is taken with respect to the escort distribution
$(p_1^{(\alpha)}, \dots, p_n^{(\alpha)})$, where $\alpha>0$ is an
additional parameter (not necessarily equal to $q$). The
interested reader is referred to our previous paper
\cite{Ilic_Stankovic_14} for this generalization.
\end{remark}

\newcommand{\od}{\op}
\newcommand{\iec}{\mc{G}}
\newcommand{\mcV}{V}
\newcommand{\uw}[1]{p_{#1}^{(q)}}
\newcommand{\mcA}[1]{{\mc G}_{#1}}

\section{On the axiomatic characterization of generalized entropy proposed in \cite{Jizba_2016}}

The entropy $\widetilde{\mc D}_n(P)$ proposed in \eqref{entropy:
jizba} does not follow generalized Shannon-Khinchin axiom [A4]
since the conditional entropy $\widetilde{\mc D}_n(Q|P)$ cannot be
represented as a quasi-linear mean \eqref{ax 4}, so that [A4] is
preserved. On the other hand, the unconditional entropy
$\widetilde{\mc D}_n(P)$ can still serve as a basis for a
generalization of statistical mechanics, as it was elaborated in
\cite{Jizba_2016}. In the following text, we propose an axiomatic
system which characterizes the entropy $\widetilde{\mc D}_n(P)$,
and give a further justification for its use.


First, recall that the entropy \eqref{entropy: jizba} can be
represented as $\widetilde{\mc D}_n(\mcP)=h(\widetilde{\mc
S}_n(P))$, where
\begin{equation}
\widetilde{\mc S}_n(P)=-\sum_{k=1}^{n} \uw{k}  \log_2 p_k
\end{equation}
is Acz\'el-Dar\'oczy entropy \cite{Aczel_Daroczy_63}. It is easy
to show that for any $P,Q \in \Delta_n$
\begin{equation}
\label{aczel entr} \widetilde{\mc S}_n(P \star Q)= \widetilde{\mc
S}_n(P)+\widetilde{\mc S}_n(Q).
\end{equation}
By applying the map $h$ to both sides of the equality \eqref{aczel
entr} and using $h(a+b)=h(a)\op_q h(b)$, we obtain
\begin{equation}
\label{aczel entr pseudoadd} \widetilde{\mc D}_n(P \star Q)=
\widetilde{\mc D}_n(P)\op_q\widetilde{\mc D}_n(Q),
\end{equation}
which means that the axiom [A4] is satisfied in the case of
independent probability distributions.

The relation \eqref{aczel entr pseudoadd} can be used as a basis
for a characterization of the entropy $\widetilde{\mc D}_n$. Let
us characterize a generalized entropy as a unique function $\mc
G_n: \Delta_n \ra \sRp$, which for all $n\in\sN$, $n>1$ satisfies
the following axioms:

\begin{description}
\item[{[C1]}]
The information content $\ec: (0,1] \ra \sR^+$ is a continuous
monotonically increasing function, which is $\op_q$-additive:
\begin{equation}
\label{Reny 1: axioms: I additivity} \ec(a b) = \ec(a) \op_q
\ec(b), \quad \text{for all} \quad a,b \in (0,1].
\end{equation}

\item[{[C2]}] $\mc G_n$ is a quasi-linear mean of the information content
\begin{equation}
\label{Reny 1: axioms: A is mean}%
\iec(\mcP) = f^{-1}\left( \sum_{k=1}^{n} \uw{k} \cdot f(\ec(p_k)
\right),
\end{equation}
for all $P \in \Delta_n$, where $f$ is invertible and continuous.%

\item[{[C3]}] $\iec_n$ is composable:
\begin{equation}
\label{Reny 1: axioms: A additivity}%
\iec_n(\mcP \star \mcQ) = \iec_n(\mcP) \op_q \iec_n(\mcQ),
\end{equation}
for all $\mcP, \mcQ \in \Delta_n$.

\item[{[C4]}] Normalization: $\iec_n\left(\frac{1}{2},\frac{1}{2}  \right) =
h(1)$.

\end{description}

Based on a result from \cite{Ilic_et_al_cert_inf} we can easily
find a class of entropies which uniquely satisfies [C1]-[C4].
According to the following theorem, the class of entropies
is even wider
than $\widetilde{\mc D}_n$, 
it depends on two parameters, $q$ and $\lambda$, and contains
$\widetilde{\mc D}_n$ as a special case for $\lambda=0$.

\begin{theorem}\label{theo: new ax sys} \rm The class of functions which uniquely satisfies [C1]-[C4] is given
by
\begin{equation}
\label{sec: iec: iec form}
\iec_n(\mcP) =%
\begin{dcases}
\isoRtoX \left(- \sum_{k=1}^{n} \uw{k} \ \log_2
p_k \right) , \quad &\lambda = 0 \\
\isoRtoX \left(\ \frac{1}{\lambda} \log_2 \left(\sum_{k=1}^{n}
\uw{k} \ p_k^{-\lambda}\right) \right), \quad &\lambda \neq 0
\end{dcases}.
\end{equation}

\end{theorem}

\textbf{Proof.} The theorem can be proven as a special case of the
theorem from \cite{Ilic_et_al_cert_inf}, where the quasi-linear
mean \eqref{Reny 1: axioms: A is mean} is not taken with respect
to the escort distribution $P^{(\alpha)}$, but with respect to
an arbitrary distribution $U$. According to Theorem 2.1. from \cite{Ilic_et_al_cert_inf}, 
the axiomatic system [C1]-[C3] is uniquely described by the
following class:
\begin{equation}
\label{sec: iec: iec form}
\iec_n(\mcP) =%
\begin{dcases}
\isoRtoX \left( \sum_{k=1}^{n} \uw{k} \ \log_2
p_k^\cop \right) , \quad &\lambda = 0 \\
\isoRtoX \left(\ \frac{1}{\lambda} \log_2 \left(\sum_{k=1}^{n}
\uw{k} \ p_k^{\cop \lambda}\right) \right), \quad &\lambda \neq 0
\end{dcases}
\quad \text{where} \quad \cop<0.
\end{equation}
If we impose the additional axiom [C4], then $\cop=-1$, and the
result follows.

\bibliographystyle{plain}
\bibliography{reference}

\begin{thebibliography}{10}

\bibitem{Abe_00}
Sumiyoshi Abe.
\newblock Axioms and uniqueness theorem for {T}sallis entropy.
\newblock {\em Physics Letters A}, 271(1--2):74 -- 79, 2000.

\bibitem{Aczel_Daroczy_63}
J.~Acz\'el and Z.~Dar\'oczy.
\newblock {{"U}ber verallgemeinerte quasilineare {M}ittelwerte, die mit
  {G}ewichtsfunktionen gebildet sind}.
\newblock {\em Publ. Math. Debrecen}, 10:171–190, 1963.

\bibitem{Hardy_et_al_34}
G.~H. Hardy, George Polya, and J.~E. Littlewood.
\newblock {\em Inequalities, by G.H. Hardy, J.E. Littlewood [and] G. Polya}.
\newblock The University press, Cambridge [Eng.], 1934.

\bibitem{Ilic_et_al_cert_inf}
Velimir~M. {Ili\'c} and Miomir~S. {Stankovi\'c}.
\newblock {A unified characterization of generalized information and certainty
  measures}.
\newblock {\em ArXiv e-prints}, October 2013.

\bibitem{Ilic_Stankovic_14}
Velimir~M. Ili\'c and Miomir~S. Stankovi\'c.
\newblock Generalized {S}hannon-{K}hinchin axioms and uniqueness theorem for
  pseudo-additive entropies.
\newblock {\em Physica A: Statistical Mechanics and its Applications}, 411:138
  -- 145, 2014.

\bibitem{Jizba_2004}
Petr Jizba and Toshihico Arimitsu.
\newblock Generalized statistics: yet another generalization.
\newblock {\em Physica A: Statistical Mechanics and its Applications},
  340(1–3):110 -- 116, 2004.
\newblock News and Expectations in Thermostatistics.

\bibitem{Jizba_2016}
Petr Jizba and Jan Korbel.
\newblock On $q$-non-extensive statistics with non-tsallisian entropy.
\newblock {\em Physica A: Statistical Mechanics and its Applications}, 444:808
  -- 827, 2016.

\bibitem{Khinchin_57}
A.~I. Khinchin.
\newblock {\em Mathematical Foundations of Information Theory}.
\newblock Dover Publications, June 1957.

\bibitem{Nivanen_Wang__03}
L.~Nivanen, A.~Le~M\'{e}haut\'{e}, and Q.~A. Wang.
\newblock {Generalized algebra within a nonextensive statistics}.
\newblock {\em Reports on Mathematical Physics}, 52(3):437--444, December 2003.

\bibitem{Tsallis_88}
Constantino Tsallis.
\newblock Possible generalization of {B}oltzmann-{G}ibbs statistics.
\newblock {\em Journal of statistical physics}, 52(1):479--487, 1988.

\end{thebibliography}

\end{document}